\documentstyle[aps,prb,preprint,epsf,epsfig,amsmath,psfrag]{revtex}
\newcommand{\beq}{\begin{equation}}
\newcommand{\eeq}{\end{equation}}

\newcommand{\inst}[1]{\mbox{$^{#1}$}}
\newcommand{\vecg}[1]{\mbox{${\boldsymbol #1}$}}

\begin{document}
\draft
\title{Domain wall roughening in dipolar films
in the presence of disorder}
\author{Jaime E. Santos\inst{1} \and Erwin Frey\inst{1} and 
Franz Schwabl\inst{2}}
\address{1 - Hahn-Meitner Institut, Abt. SF5, Glienicker Str. 100,
14109 Berlin, Germany\\
2 - Institut f\"ur Theoretische Physik T34, 
Physik-Department der 
TU M\"unchen,
James-Franck-Stra\ss e, 85747 Garching, Germany\\
email: {\rm santos@hmi.de} \\
email: {\rm frey@hmi.de}\\
email: {\rm schwabl@physik.tu-muenchen.de}
} 
\maketitle
\date{}
\newpage
\begin{abstract}
We derive a low-energy Hamiltonian for the elastic energy of
a N\'eel domain wall in a thin film with in-plane magnetization,
where we consider the contribution of the long-range dipolar
interaction beyond the quadratic approximation. 
We show that such a Hamiltonian is analogous to the
Hamiltonian of a one-dimensional polaron in an external random
potential. We use a replica variational method to compute the
roughening exponent of the domain wall for the case of two-dimensional
dipolar interactions.
\end{abstract}
\pacs{05.70.Np, 68.55.Ln}
\newpage
\section{Introduction}
\label{secA}
The roughening properties of elastic manifolds in the
presence of quenched disorder is a well studied problem. 
The subject of study is usually the asymptotic
(long-distance) regime in which one wishes to compute
the roughening exponent $\zeta$ of the $D$-dimensional
elastic manifold (described by a displacement
field $\phi(\vecg{x})$)
defined by
\begin{equation} 
\overline{\langle (\phi(\vecg{x})-\phi(\vecg{x}'))^2\rangle}=
\mid \vecg{x}-\vecg{x}'\mid^{2\zeta},
\label{eq-1}
\end{equation}
where $\vecg{x}$ is the position vector in the $D$-dimensional
configuration space and where the brackets represent an average
over thermal fluctuations and the overline an average
over the realizations of the
quenched disorder. Obviously, the study of the crossover
between short distances, where the thermal fluctuations dominate
and the long-distance behaviour, which is determined by the disorder,
is also of interest.
Among the methods applied to the study of this problem are 
the use of Imry-Ma type of arguments \cite{Grinstein82,Villain82,Imry75}, 
the mapping
of one-dimensional interfaces in two dimensions to the noisy Burgers
equation \cite{Burgers74}, which yields an exact result
for the roughening exponent
\cite{Huse85,Zhang86}, variational approaches involving 
replica averaging and replica symmetry breaking
\cite{Mezard90,Gold00,Haza99} and functional renormalization
group calculations \cite{Balents93,Emig98}. The role
played by long-range interactions does not seem to
be so well understood, although it follows
from the one-loop functional
RG results of Emig and Nattermann \cite{Emig98} 
that in the important case of a magnetic domain-wall 
in the presence of long-range dipolar interactions, one simply
needs to replace the expansion parameter $\epsilon=4-D$
by $\epsilon=3(3-D)/2$ (where $D$ is the dimension
of the manifold) in the expressions
for the critical exponents obtained in
the absence of dipolar interactions,
in order to account for the presence of these interactions
\cite{Kolomeiskii88}.
In particular, if one applies this result
to the case of a line domain in a thin 
magnetic film ($D=1$), the effective expansion parameter is
$\epsilon=3$ in both cases, which means that in this
case and to one-loop order, the critical exponents 
obtained are the same in the absence or presence of
dipolar interactions.
Although one expects the RG calculations to
provide a qualitative understanding at low
dimensions compared to the upper critical dimension
($D=3$ in this case), it is still questionable to what
extend one can perform such analytical continuation using
just the one-loop results. Furthermore, the results of 
Emig and Nattermann were based on a series expansion 
of the dipolar energy \cite{Lajz80,Natterman83} 
which meant that, in  the absence of disorder, 
the Hamiltonian describing the low energy degrees of freedom of
the domain wall was quadratic in the domain-wall
displacement field, i.e. the random potential
is the only source of non-linearity in the problem.
One still needs to justify that such approximation 
is sufficient \cite{Note0}.

Having in mind such difficulties, we wish to
discuss the properties of a one-dimensional domain wall
in a two-dimensional ferromagnetic film, in the presence of dipolar
interactions and a short-range correlated random field,
where the dipolar interaction is treated beyond the
quadratic approximation. We wish to check whether
it is still possible to map such a problem to a
(modified) Burgers equation, as was done by Huse et al.
\cite{Huse85} for the case of random-bond disorder
and by Zhang \cite{Zhang86} for the case of random-field disorder,
in the absence of long-range interactions. Such a mapping
was achieved because one can map the line domain to the
world line of a quantum particle in 1+1 dimensions. The
free energy of the line domain can then be shown to obey
a Schr\"odinger equation in imaginary time in the presence
of a random potential. A Cole-Hopf transformation can then
be used to convert this equation into a noisy Burgers equation
with additive noise, for which the stationary distribution
is known in one-dimension \cite{Deker75}. This allows one
to compute the roughening exponent. In our
case, the presence of the dipolar interaction does not
allow one to write the energy of the domain wall as an action
for a single quantum particle, but instead we obtain an
action for a quantum particle which interacts with a 
annealed `vector potential', whose 
propagator is determined by the dipolar propagator. In 1+1 
dimensions, a gauge transformation allows us to 
substitute the vector potential
by a scalar one, which can then be integrated out, 
leaving us with
an action similar to the action of a polaron 
as considered by Feynman \cite{Feynman56,Spohn86}, 
i.e. the dipolar interaction introduces 
a `self-retarded' interaction of the particle with itself.
Despite the fact that such action cannot be integrated exactly,
we can still apply a variational method in the spirit of
Feynman \cite{Mezard90,Gold00,Haza99}, in which one introduces
replicas of the system in order to average the free energy
of the domain wall over the realizations of the random field.
Using the hierarchical replica symmetry breaking ansatz of Parisi
\cite{Parisi80}, we were able to derive a set of non-linear
equations whose self-consistent solution yields
the long-distance physics of the problem
within the realm of this approximation. 
With this method, we are able to treat two different
situations. The one of a thin ferromagnetic 
film grown on a non-magnetic
substrate, in which the dipolar interaction has the usual
three dimensional form, and the one of a thin
ferromagnetic film grown between two type I superconductors,
i.e. the case of a superconductor/ferromagnet/superconductor (SC/FM/SC)
heterostructure. We have shown elsewhere \cite{Santos00}
that in this case the dipolar interaction has a two-dimensional
form at small wave-vectors, coming from the renormalization
of the magnetic energy due to the Meissner effect in the
superconductors. Although we were unable to obtain
an analytic solution of the variational non-linear equations, 
we were able to justify that in the case of 
two-dimensional dipolar interactions (i.e. for the SC/FM/SC heterostructure), 
the quadratic approximation is indeed adequate.
This case is not only easier to treat analytically, it is
also more interesting than the case of three dimensional
dipolar interactions, for which one expects the dipolar
interaction to be irrelevant (see above). 
Therefore, we restrict our discussion
to the case of two-dimensional dipolar interactions
in this paper. Furthermore, we can also show 
that in this case, if one works within the quadratic approximation, 
the elastic energy of the domain-wall has the same analytical form 
as the energy of a domain wall in a 
ferroelastic material as studied by Kolomeiskii et al.
\cite{Kolomeiskii88} and also of the energy  
of a liquid-gas interface in a disordered solid, as studied
by Hazareesing and M\'ezard \cite{Haza99}. 
In both cases, one obtains, in two dimensions,
a value of $1/3$ for the roughening exponent.
This is indeed the value that we obtain for the
roughening exponent in our problem. It should be 
nevertheless noticed that these
three problems, though mathematically similar have a
different physical origin and motivation.

The structure of this paper is as follows: in section II, we derive
a low-energy Hamiltonian describing the N\'eel domain-wall
elastic degrees of freedom. This Hamiltonian can be 
interpreted as the action in imaginary time of a 
quantum particle in 1+1 dimensions. 
In section III, we use the replica trick to average over the random
field realizations, which generates a many particle non-quadratic
action. In section IV, we introduce a generalization
of the variational ansatz of M\'ezard and Parisi 
which allows us to consider the effect of the short-range
correlated disorder and of the non-local dipolar term in the many
particle action. One obtains a set
of non-linear equations which have to be solved self-consistently.
In section V, we perform the analysis of these equations
in the particular case of the hierarchical replica symmetry 
breaking solution of Parisi and we
discuss the validity of the quadratic approximation for the dipolar
term in these equations, which permits to determine the roughening
exponent of the domain wall. Finally, in section VI, we present
our conclusions.

\section{Energy of a single distorted domain wall in a 
two-dimensional XY model with 
dipolar interactions and in a random field}
\label{secB}
We consider a N\'eel (i.e. $180$ degrees) domain-wall in a
thin two-dimensional ferromagnetic
film in the presence of dipolar interactions and a short-range
correlated quenched random field. The magnetization is in the plane
of the film. In the straight domain confi\-gu\-ra\-tion,
the wall is oriented along the $y$ axis and the center of
the wall is located at $x=0$. The thermal and random field
fluctuations give rise to deviations of the domain-wall 
from its straight configuration,
i.e. at a given position $y$, the center of the domain is
displaced from $x=0$ to $x(y)$ (we neglect overhangs).
In this case, equation (\ref{eq-1}) reduces to
\begin{equation} 
\overline{\langle (x(y)-x(y'))^2\rangle}=\mid y-y'\mid^{2\zeta}\,.
\label{eq0}
\end{equation}
The distance $\mid y-y'\mid$
is considered to be very large with respect to the lattice
spacing and to the length over which the random field
is correlated (see below).
In the absence of dipolar interactions, one can,
as stated above, through
a mapping to a Schr\"odinger equation 
\cite{Zhang86},
show that $\zeta=1$. 

In order to determine the elastic energy of the
domain-wall, we consider the underlying system of 
ferromagnetic spins to be in a square
lattice with lattice constant $a$ and overall
dimensions $L_x\times L_y$. The system has 
anti-periodic boundary conditions along the $x$ direction 
(in order that there is a single domain-wall in the system)
and periodic boundary conditions
along the $y$ direction and is at finite temperature $T$.
Such system is described
by a (classical) XY model Hamiltonian,
with dipolar interactions and in a random field,
\begin{eqnarray}
\label{eq1}
{\cal H}&=&-\,\frac{1}{2}\sum_{<i,j>}\,J_{ij}\,\vecg{S}_i\cdot\vecg{S}_{j}
-\mu\,\sum_{i}\,\vecg{h}_i\cdot\vecg{S}_i\\
& &\mbox{}+\frac{G}{2}\,\sum_{i\neq j}\left(
\frac{\vecg{S}_i\cdot\vecg{S}_{j}}{\mid \vecg{R}_i -\vecg{R}_j\mid^n}
-\frac{n\,(\vecg{R}_i-\vecg{R}_j)\cdot\vecg{S}_i\,
(\vecg{R}_i-\vecg{R}_j)\cdot\vecg{S}_j}{
\mid \vecg{R}_i -\vecg{R}_j\mid^{n+2}}\right)
-D\,\sum_{i}\,(S_i^y)^2\,.\nonumber
\end{eqnarray}
For the case of two-dimensional dipolar interactions
(i.e. for the SC/FM/SC heterostructure),
the constants $\mu$ and $G$ are given in the MKS system
by $\mu=\mu_0\,\mu_B\,g_L$ and
$G=\frac{\mu_0}{4\pi\mu_r\lambda_L}(g_L\mu_B)^2$,
where $\mu_0$ is 
the magnetic permitivity of the vacuum, $\mu_B$
the Bohr magneton and $g_L$ the Land\'e factor of the spin
system. The constants $\lambda_L$
and $\mu_r$ are, respectively, the London penetration depth and 
relative magnetic permitivity of the superconductor.
In this case, $n=2$. In the case of three dimensional dipolar
interactions, which we will not consider,
$G=\frac{\mu_0}{4\pi}(\mu_B\,g_L)^2$ and $n$ is equal to $3$.
The quantities $J_{ij}$ and $D$ have the dimensions of
energy \cite{NoteMKS}. 
The field $\vecg{h}_i$ is a random field with Gaussian
correlations, i.e. 
$\overline{h_i^\alpha\,h_j^\beta}=\Delta_{ij}\delta^{\alpha\beta}$. 
In order to stabilize the domain-wall, 
we have added  to the Hamiltonian
above an easy-axis anisotropy term, along the $y$ axis,
represented by the last term in (\ref{eq1}).
 
For a straight N\'eel domain-wall, the spin configuration
has the form
\begin{eqnarray}
\label{eq2}
S_i^x&=&S\,\cos\phi(x_i)\\
S_i^y&=&S\,\sin\phi(x_i)
\label{eq3}
\end{eqnarray}
where $i$ is the site label
and where $\phi(x)$ goes from  
$-\pi/2$ when $x\rightarrow -L_x/2$ to
$\pi/2$ when $x\rightarrow L_x/2$, 
with $S$ being the value
of the spin. The function $\phi(x)$ changes 
its value from $-\pi/2$ to $\pi/2$ in a region of
width $w$ around $x=0$. Since $w$ is typically of the order 
of tens of nanometers \cite{OA} and we are only interested
in the long-distance properties of the domain-wall,
we can write $\frac{d\phi(x)}{dx}=\pi\,\delta(x)$.
If the center of the domain-wall is displaced from
$x_i=0$ to $x_i=x(y_i)$, the spin configuration becomes
\begin{eqnarray}
\label{eq4}
S_i^x&=&S\,\cos\phi(x_i-x(y_i))\\
S_i^y&=&S\,\sin\phi(x_i-x(y_i))\, .
\label{eq5}
\end{eqnarray}
We wish to compute the partition function
${\cal Z}=Tr(e^{-\delta{\cal H}/k_B T})$, where 
$\delta{\cal H}=\delta{\cal H}_{exc}+\delta{\cal H}_{rf}
+\delta{\cal H}_{dip}$ is the energy difference between
a configuration described by the displacement 
$x(y)$ and the straight domain configuration.
We define the discrete Fourier transforms
\begin{eqnarray}
\label{eq6}
S_{\vecg{q}}^\alpha
&=&\sum_{\vecg{R}_i}\,S_i^\alpha\,e^{-i\vecg{q}\cdot\vecg{R}_i}\\
x(q_y)&=&\sum_{y_i}\,x(y_i)\,e^{-iq_y y_i}
\label{eq7}
\end{eqnarray}
where $\vecg{q}$ belongs to the first Brillouin zone.
We will approximate the sums over $i$ by integrals,
according to the formulas,
\begin{equation}
\sum_{\vecg{R}_i}
\approx\frac{1}{a^2}\int\,d^2x\;\;,\;\; \sum_{y_i}\approx\frac{1}{a}
\int\,dy\, ,
\label{eq8}
\end{equation}
valid for a square lattice. It can be shown that 
the (positive) contribution to the exchange energy
coming from the domain-wall deformation is given
to second order in the displacement field, by \cite{Lajz80}
\begin{eqnarray}
\delta{\cal H}_{exc}&=&
-\frac{\pi^2S^2}{2Na^4}\sum_{\vecg{q}}(J(q_x,q_y)-J(q_x,0))\,
x(q_y)\,x(-q_y)\nonumber\\
&\approx&\frac{\pi^2\,J\,S^2}{2L_ya}\sum_{q_y}q_y^2\,x(q_y)\,
x(-q_y)
\label{eq9}
\end{eqnarray}
where $N=L_xL_y/a^2$ is the total number of sites and
where we have expanded $J(\vecg{q})\approx J_0 -Ja^2q^2$.
In real space, equation (\ref{eq9}) reads
\begin{eqnarray}
\delta{\cal H}_{exc}&=&
\frac{1}{2}\,\sigma\,
\int_{-\frac{L_y}{2}}^{\frac{L_y}{2}}\, 
dy\,\left(\frac{dx}{dy}\right)^2\, ,
\label{eq10}
\end{eqnarray}
which has the same form as the kinetic part of the action
of a particle of mass $\sigma$ in one dimension, with
$\sigma=\pi^2JS^2/a$ being the line-tension of the domain
wall.
The contribution coming from the random field can be shown
to be equal to \cite{Random?}
\begin{equation}
\delta{\cal H}_{rf}=\tilde{\mu}\,
\int_{-\frac{L_y}{2}}^{\frac{L_y}{2}}\, dy\,
\int_{0}^{x(y)}\,dx\,\tilde{h}^y(x,y)
\label{eq11}
\end{equation}
where $\tilde{\mu}=\mu_0\,\mu_B\,g_L\pi S/a$ and $\tilde{h}^y=h^y/a$
\cite{Notesrf}.
This formula can be made plausible if we consider instead a constant
field along $y$. It can also be easily seen from (\ref{eq9}) 
that the easy-axis anisotropy
term does not contribute to the deformation 
energy of the domain-wall.

Since we wish to compute the partition function,
we have to consider the factor  $e^{-{\cal H}_{dip}/k_B T}$, 
where ${\cal H}_{dip}$ is given by
\begin{equation}
{\cal H}_{dip}=\frac{g}{2V}\, \sum_{\vecg{q}}\,{\cal F}(\vecg{q})
\,(\vecg{q}\cdot\vecg{S}_{\vecg{q}})\,(\vecg{q}\cdot\vecg{S}_{-\vecg{q}})\, ,
\label{eq12}
\end{equation}
where $V=L_xL_y$ is the area of the sample. 
For two-dimensional dipolar interactions, 
$g=\frac{\mu_0}{2\mu_r\lambda_L}(g_L\mu_B)^2$ and 
${\cal F}(\vecg{q})=\frac{1}{q^2}$ \cite{Santos00}.
For three dimensional dipolar interactions, 
$g=\frac{\mu_0}{2}(g_L\mu_B)^2$ 
and ${\cal F}(\vecg{q})=\frac{1}{q}$.
Thus, by considering the different dipolar kernels
${\cal F}(\vecg{q})$, one can treat the two different 
cases. 

Using a Hubbard-Stratonovich transformation, we write
$e^{-{\cal H}_{dip}/k_B T}$ as
\begin{equation}
e^{-\frac{{\cal H}_{dip}}{k_B T}}={\cal N}\,
\int\,{\cal D}A_{\vecg{q}}\,\exp\left(\,-\frac{1}{2Vk_B T}\sum_{\vecg{q}}
\, A_{\vecg{q}}\,{\cal F}^{-1}(\vecg{q})\,A_{-\vecg{q}}\,+\,
\frac{\sqrt{g}}{Vk_B T}\sum_{\vecg{q}}
\, (\vecg{q}\cdot\vecg{S}_{\vecg{q}})\,A_{-\vecg{q}}\,\right)
\label{eq13}
\end{equation}
where ${\cal N}$ is a normalization factor.
We can now show that
\begin{equation}
\vecg{q}\cdot\vecg{S}_{\vecg{q}}=\frac{i\pi S}{a^2}\,
\int^{\frac{L_y}{2}}_{-\frac{L_y}{2}}\,dy\,e^{-iq_x x(y)-iq_y y}\,
\frac{dx}{dy}\,,
\label{eq14}
\end{equation}
from which it follows that
{\small
\begin{equation}
\frac{1}{V}
\sum_{\vecg{q}}(\vecg{q}\cdot\vecg{S}_{\vecg{q}})A_{-\vecg{q}}
=\frac{i\pi S}{a^2}
\int^{\frac{L_y}{2}}_{-\frac{L_y}{2}}\,dy\,A(x(y),y)\,\frac{dx}{dy}
\label{eq15}
\end{equation}}
and one can see that there is no contribution to the dipolar
energy when $x(y)=const$. The partition function 
of the domain wall's degrees of freedom involves 
the functional integration of the Boltzmann
factor $e^{-\delta{\cal H}/k_B T}$, over the space of functions $x(y)$.

We shall now make a change in our notation. We will write
\begin{eqnarray}
&&y\rightarrow \tau,\; x(y)\rightarrow x(\tau),\;
\frac{dx}{dy}\rightarrow \dot{x}(\tau),\nonumber\\
&&q_y\rightarrow \omega_m,\;
e=\frac{\pi S\sqrt{g}}{a^2}\,.
\label{eq16}
\end{eqnarray}
The subscript $m$ stands for the fact that the
wave-vectors $\omega_m$ are discrete,
due to the periodic boundary conditions along $\tau$ ($y$).
If we use this notation
in the formulae above, we have, 
collecting all the factors, the following expression
for the partition function
\begin{equation}
{\cal Z}={\cal N}\,\int\,{\cal D}x(\tau)\,{\cal D}A(q_x,\omega_m)
\,e^{-{S}/k_B T}
\label{eq17}
\end{equation}
where the action $S$ is given by
\begin{eqnarray}
\label{eq18}
S&=&\frac{1}{2}\,\sigma
\int_{-\frac{L_y}{2}}^{\frac{L_y}{2}}\, 
d\tau\,\dot{x}^2(\tau)+\tilde{\mu}\,
\int_{-\frac{L_y}{2}}^{\frac{L_y}{2}}\, d\tau\,
\int_{0}^{x(\tau)}\,dx\,\tilde{h}^y(x,\tau)\nonumber\\
&&\mbox{}-ie
\int^{\frac{L_y}{2}}_{-\frac{L_y}{2}}\,
d\tau\,A(x(\tau),\tau)\,\dot{x}(\tau)
+\frac{1}{2V}\sum_{q_x,\omega_m}
\, A(q_x,\omega_m)\,{\cal F}^{-1}(q_x,\omega_m)\,A(-q_x,-\omega_m)\,.
\end{eqnarray}
This is the action of a single quantum particle in 1+1
dimensions, with `mass' $\sigma$ and `charge' $e$, 
at `inverse temperature' $L_y$
and with a `Planck constant' equal to $k_B T$. 
Such  particle is in a external random potential 
$V(x,\tau)=\tilde{\mu}\,
\int_{0}^{x}\,dx'\,\tilde{h}^y(x',\tau)$ and interacts with
a `vector potential' $A(q_x,\omega_m)$ \cite{Kleinert?} 
characterized by a
propagator equal to ${\cal F}(q_x,\omega_m)$.
This result belongs to the well known 
class of mappings of $2$ dimensional problems
of classical statistical mechanics
to $1+1$ dimensional quantum problems.
Now, since one has periodic boundary conditions 
along the $\tau$ ($y$) axis, the following identity
holds
\begin{equation}
\int^{\frac{L_y}{2}}_{-\frac{L_y}{2}}
d\tau\,A(x(\tau),\tau)\,\dot{x}(\tau)=
-\int^{\frac{L_y}{2}}_{-\frac{L_y}{2}}
d\tau\int_{0}^{x(\tau)}dx'\,
\frac{\partial A(x',\tau)}{\partial\tau}\,,
\label{eq19}
\end{equation}
which corresponds to a gauge transformation of the action
\cite{NoteG}.
The use of this identity eliminates the dependence on the
`velocity' $\dot{x}(\tau)$ in all terms of equation
(\ref{eq18}) except the first one. 
The action $S$ now reads
\begin{eqnarray}
\label{eq20}
S&=&\frac{1}{2}\,\sigma
\int_{-\frac{L_y}{2}}^{\frac{L_y}{2}}\, 
d\tau\,\dot{x}^2(\tau)+\,\int_{-\frac{L_y}{2}}^{
\frac{L_y}{2}}\, d\tau\,V(x(\tau),\tau)
+ie\int^{\frac{L_y}{2}}_{-\frac{L_y}{2}}\,
d\tau\,\int_{0}^{x(\tau)}dx'\,
\frac{\partial A(x',\tau)}{\partial\tau}\nonumber\\
&&\mbox{}+\frac{1}{2V}\sum_{q_x,\omega_m}
\,A(q_x,\omega_m)\,{\cal F}^{-1}(q_x,\omega_m)\,A(-q_x,-\omega_m)\,.
\end{eqnarray}
We can now integrate out the field $A(q_x,\omega_m)$, 
obtaining an action entirely in terms of $x(\tau)$ alone. We obtain
\begin{eqnarray}
\label{eq21}
S&=&\frac{1}{2}\,\sigma
\int_{-\frac{L_y}{2}}^{\frac{L_y}{2}}\, 
d\tau\,\dot{x}^2(\tau)+\,\int_{-\frac{L_y}{2}}^{
\frac{L_y}{2}}\, d\tau\,V(x(\tau),\tau)\nonumber\\
&&\mbox{}+\frac{e^2}{2V}\,\sum_{q_x,\omega_m}\,
\frac{\omega_m^2}{q_x^2}\,{\cal F}(q_x,\omega_m)\,
\int_{-\frac{L_y}{2}}^{\frac{L_y}{2}}\, d\tau\,
\int_{-\frac{L_y}{2}}^{\frac{L_y}{2}}\, d\tau'\,
e^{-iq_x(x(\tau)-x(\tau'))-i\omega_m(\tau-\tau')}\, .
\end{eqnarray}
This action can be used to compute the roughening
exponent of the domain wall. Equation (\ref{eq21})
also shows that, apart from the interaction with the $\tau$ dependent
random potential $V(x,\tau)$, the action $S$ is
analogous to the action of a one-dimensional polaron,
as written by Feynman \cite{Feynman56,Spohn86}. 

However, in order to compute the roughening
exponent, one has to determine the following quantity
\begin{equation}
\overline{\langle (x(\tau)-x(\tau'))^2\rangle}=
\overline{\frac{\int\,{\cal D}x(\tau)\, e^{-S/k_BT}\,
(x(\tau)-x(\tau'))^2}{\int\,{\cal D}x(\tau)\, e^{-S/k_BT}}}
\label{eq23}
\end{equation}
which involves determining the average of the
distribution of quenched disorder over a quotient.
This type of averaging places the same problem as 
when one needs to average over the free energy 
of the system, i.e. one needs to determine the average
$\overline{\log {\cal Z}}$, instead of
simply averaging over the partition function ${\cal Z}$. 
This can be seen by including a source field coupled to
$x(\tau)$ in ${\cal Z}$. The two-point correlation
function appearing in (\ref{eq23}) can be computed from 
$\overline{\log {\cal Z}}$ by differentiating
it with respect to the sources and taking these
sources to be zero. 
\section{The average over disorder. Replica trick.}
\label{secC}
We need to perform the average over the random
field $\tilde{h}^y(q_x,\omega_m)$, which obeys a Gaussian distribution 
with a second moment equal to
\begin{equation}
\langle \tilde{h}^y(q_x,\omega_m) \tilde{h}^y(q_x',\omega_m')\rangle
=\frac{N}{a^2}\,\Delta(q_x,\omega_m)\,\delta_{q_x+q_x',0}
\,\delta_{\omega_m+\omega_m',0}.
\label{eq24}
\end{equation}
In order to perform the averaging over disorder in the
free energy, one introduces $n$ distinct copies (replicas)
of the system \cite{Edwards75} and makes use of the
identity
\begin{equation}
\log{\cal Z}=\lim_{n\rightarrow 0}
\frac{{\cal Z}^n -1}{n}.
\label{eq25}
\end{equation}
One performs the disorder average over ${\cal Z}^n$ for integer $n$
and makes the analytical continuation to $n\rightarrow 0$.
One obtains 
\begin{equation}
\overline{{\cal Z}^n}=\int\,\prod_a {\cal D}\,
x_a(\tau)\,e^{-S_R/k_B T},
\label{eq26}
\end{equation} 
where the `replicated' action $S_R$ is given by
\begin{eqnarray}
\label{eq27}
S_R&=&\frac{1}{2}\,\sigma\,\sum_a\,
\int_{-\frac{L_y}{2}}^{\frac{L_y}{2}}\, 
d\tau\,(\,\dot{x}_a^2(\tau)+ \Omega_0^2\,x_a^2(\tau)\,)
+\frac{\tilde{\mu}^2n}{k_B TL_x}\sum_{a,q_x}
\frac{\Delta(q_x,0)}{q_x^2}\,
\int_{-\frac{L_y}{2}}^{
\frac{L_y}{2}}\, d\tau\,\cos(q_x x_a(\tau))\nonumber\\
&&\mbox{}+\frac{e^2}{2V}\,\sum_{a,q_x,\omega_m}\,
\frac{\omega_m^2}{q_x^2}\,{\cal F}(q_x,\omega_m)
\int_{-\frac{L_y}{2}}^{\frac{L_y}{2}}d\tau
\int_{-\frac{L_y}{2}}^{\frac{L_y}{2}}d\tau'
e^{-iq_x(x_a(\tau)-x_a(\tau'))-i\omega_m(\tau-\tau')}\nonumber\\
&&\mbox{}-\frac{\tilde{\mu}^2}{k_BTV}\sum_{a,b,q_x,\omega_m}
\frac{\Delta(q_x,\omega_m)}{q_x^2}\,
\int_{-\frac{L_y}{2}}^{\frac{L_y}{2}}d\tau
\int_{-\frac{L_y}{2}}^{\frac{L_y}{2}}d\tau'
e^{-iq_x(x_a(\tau)-x_b(\tau'))-i\omega_m(\tau-\tau')}
\end{eqnarray}
and where the last term represents the inter-replica interaction,
due to the presence of the random potential.
We have included 
in $S_R$ the
effect of an applied field $\tilde{h}^y_{ext}(x)=
(\sigma\Omega_0^2/\tilde{\mu})\,x$. This term gives
rise to the second term in the action, which is
a harmonic potential.
The presence of this term is necessary
to guarantee that $S_R$ is bounded from below,
given that the last term of (\ref{eq27})
comes with a negative sign. $\Omega_0$ can 
be set to $0$ at the end of the calculations.
\section{A variational ansatz for the free energy. 
Variational equations.}
\label{secD}
The action (\ref{eq27}) cannot be integrated exactly. Therefore,
we need to develop an approximation scheme. We shall follow
M\'{e}zard and Parisi \cite{Mezard90} and Goldschmidt \cite{Gold00}. 
We choose a trial quadratic action
of the form
\begin{eqnarray}
S_0[\eta_a(\tau)]&=&\frac{1}{2}\,\sigma\,\sum_a\,
\int_{-\frac{L_y}{2}}^{\frac{L_y}{2}}\, 
d\tau\,(\,\dot{x}_a^2(\tau)+ \Omega_a^2\,x_a^2(\tau)\,)\nonumber\\
&&\mbox{}-\frac{1}{2}\sum_{a,b}
\int_{-\frac{L_y}{2}}^{\frac{L_y}{2}}d\tau
\int_{-\frac{L_y}{2}}^{\frac{L_y}{2}}d\tau'\,
x_a(\tau)\,Q_{ab}(\tau-\tau')\,x_b(\tau')
\nonumber\\
&&\mbox{}-
\sum_{a}\int_{-\frac{L_y}{2}}^{\frac{L_y}{2}}d\tau\,
\eta_a(\tau)\,x_a(\tau)
\label{eq28}
\end{eqnarray}
where we have included a source field $\eta_a(\tau)$, which
is useful in the calculation of correlation functions.
The M\'{e}zard-Parisi ansatz has been generalized by
including a matrix function $Q_{ab}(\tau-\tau')$, with a non-trivial
dependence on $\tau-\tau'$,
necessary to take into account the effect
of the non-local dipolar interaction. In the
M\'ezard and Parisi case, $Q_{ab}(\tau-\tau')=
Q_{ab}\,\delta(\tau-\tau')$,
since they considered only short-range correlated disorder.
Goldschmidt has considered several different forms for the kernel
$Q_{ab}(\tau-\tau')$,  all of which non-local in $\tau-\tau'$,
and which are able to account for the problem 
of a quantum particle in a disordered
static potential and for the related problem of a polymer in a random
potential with long-range correlated disorder (see below). 
Since $Q_{ab}(\tau)=Q_{ab}(\tau+L_y)$, 
we can represent $Q_{ab}(\tau)$ in terms of Matsubara
modes $Q_{ab}(\tau)=1/L_y \sum_m Q_{ab}^m e^{i\omega_m \tau}$.
The values of $\Omega_a^2$ and $Q_{ab}^m$ 
are to be chosen appropriately. 
We also demand $S_0$ to be translationally
invariant when $\Omega_a\rightarrow 0$. This gives rise
to the constraint 
\begin{equation}
Q_{bb}^0=-\sum_{a\neq b}Q_{ab}^0\,.
\label{eq29}
\end{equation}
It can be shown that the normalized generating
function for the correlation functions of such
a quadratic action, is given by \cite{Ada82}
\begin{equation}
Z_N[\eta_a(\tau)]=
\exp\left(1/2(k_B T)^2\sum_{ab}
\int_{-\frac{L_y}{2}}^{\frac{L_y}{2}}d\tau 
\int_{-\frac{L_y}{2}}^{\frac{L_y}{2}}d\tau'\,
\eta_a(\tau)\,G_{ab}(\tau-\tau')\,\eta_b(\tau')\right)
\label{eq30}
\end{equation}
where $G_{ab}(\tau-\tau')=\langle x_a(\tau)x_b(\tau')\rangle$
is given in terms of its Matsubara modes, by
\begin{equation}
G_{ab}^m=k_B T\,
[\,\sigma(\omega_m^2\,\hat{I}+\hat{\Omega}^2)-\hat{Q}^m\,]^{-1}_{ab},
\label{eq31}
\end{equation}
where $\hat{I}$ is the unit matrix in replica space and $\hat{\Omega}^2$
is the diagonal matrix with elements equal to $\Omega_a^2$.
The free energy as computed from the trial action $S_0[0]$,
with the sources $\eta_a(\tau)$ equal to zero, can
be shown to be equal to
\begin{eqnarray}
F_0&=&C^t
-\frac{k_BT}{2}\sum_{m,\mu}\ln\,[\,(1/k_BT)(\sigma\omega_m^2
+(\sigma\hat{\Omega}^2-\hat{Q}^m)_\mu)\,]
\label{eq32}
\end{eqnarray}
where $\mu$ indicates the basis in which the matrix
$\sigma\hat{\Omega}^2-\hat{Q}^m$ is diagonal.
One now uses the Feynman inequality 
for $F=-k_B T\log\overline{{\cal Z}^n}$, which states that,
\begin{equation}
F\leq F_{var}=F_0 +\langle\, S_R -S_0[0]\,\rangle_0,
\label{eq33}
\end{equation}
where the average in the rhs of (\ref{eq33}) is over the
trial quadratic action $S_0[0]$. Note that 
$ {\cal F}= -k_B T \,\overline{\log {\cal Z}}= \lim_{n\rightarrow 0}
F/n$. One obtains for $F_{var}$ the expression
\begin{eqnarray}
\label{eq34}
\frac{F_{var}}{L_y}&=&\tilde{C}^t
-\frac{k_BT}{2L_y}\sum_{m,\mu}\ln\,
[\,(1/k_BT)(\sigma\omega_m^2
+(\sigma\hat{\Omega}^2-\hat{Q}^m)_\mu)\,] \\
&&\mbox{}+\frac{\sigma}{2 L_y}\sum_{a,m}
(\Omega_0^2 -\Omega_a^2)
[\,(1/k_BT)(\sigma\omega_m^2\hat{I}
+(\sigma\hat{\Omega}^2-\hat{Q}^m))\,]_{aa}\nonumber\\
&&\mbox{}+\frac{1}{2 L_y}\sum_{a,b,m}Q_{ab}^m\,
[\,(1/k_BT)(\sigma\omega_m^2\hat{I}
+(\sigma\hat{\Omega}^2-\hat{Q}^m))\,]_{ba}\nonumber\\
&&\mbox{}+\frac{\tilde{\mu}^2n}{k_B T L_x}\sum_{a,q_x}
\frac{\Delta(q_x,0)}{q_x^2}\,e^{-\frac{1}{2}\,q_x^2\,G_{aa}(0)}
+\frac{1}{2V}\sum_{a,b,q_x,\omega}
\int_{-\frac{L_y}{2}}^{\frac{L_y}{2}}d\tau
\left(\frac{e^2\omega^2}{q_x^2}\,{\cal F}(q_x,\omega)\,\delta_{ab}
\right.\nonumber\\
&&\mbox{}-\left.
\frac{\tilde{\mu}^2}{k_B T}\frac{\Delta(q_x,\omega)}{q_x^2}
\right)e^{-i\omega\tau-\frac{1}{2}\,q_x^2\,(
G_{aa}(0)+G_{bb}(0)-2G_{ab}(\tau))}\nonumber
\end{eqnarray}
where the following identities, valid for a quadratic
action at zero sources, 
$\langle e^{-iq_x x_a(\tau)}\rangle_0
=e^{-\frac{1}{2}\,q_x^2\,G_{aa}(0)}$ 
and $\langle e^{-iq_x(x_a(\tau)
-x_b(0))}\rangle_0=e^{-\frac{1}{2}\,q_x^2\,(
G_{aa}(0)+G_{bb}(0)-2G_{ab}(\tau))}$ were used.
Following Hazareesing and M\'ezard \cite{Haza99},
we choose the random field to have
anisotropic correlations,
$\Delta(q_x,\omega_m)=\Delta\,e^{-1/2\,q^2_x\,{\cal B}^2}$,
with correlation length ${\cal B}$ along $x$ \cite{NoteCorr,NoteGold_1}
and we choose the two-dimensional form of the dipolar kernel,
i.e. ${\cal F}(q_x,\omega_m)=1/(q_x^2+\omega_m^2)$.
We have to minimize (\ref{eq34}) with respect
to the free parameters $Q_{ab}^m$ and $\Omega_a^2$.
Such procedure gives the following
self-consistent equations
\begin{eqnarray}
\label{eq35}
\Omega_a^2&=&
\Omega_0^2-\frac{\tilde{\mu}^2\,\Delta\,n}{\sqrt{2\pi}k_B T}
({\cal B}^2+G_{aa}(0))^{-1/2}\\
\label{eq36}
Q_{ab}^m &=& Q_{ab}^0+f_a^m\delta_{ab}\;\;m\neq 0\\
&&\mbox{with}\nonumber\\
\label{eq37}
Q_{ab}^0&=&
\frac{\tilde{\mu}^2\Delta}{\sqrt{2\pi}k_B T}
(\,{\cal B}^2 +G_{aa}(0)+G_{bb}(0)-2G_{ab}(0)\,)^{-1/2}\\
&& (a\neq b)\nonumber\\
\label{eq38}
f^m_a&=&
-e^2\int_0^{L_y/2}\,d\tau
(\,1-\cos(\omega_m\tau)\,)\,\left[\,\frac{1}{2\pi \gamma_{aa}(\tau)}
\right.\nonumber\\
&&\mbox{}\left.
-\frac{\tau}{4\,\sqrt{\pi}\,\gamma_{aa}^{3/2}(\tau)}\,
e^{\frac{\tau^2}{4\gamma_{aa}(\tau)}}\,
\mbox{Erfc}\left(\frac{\tau}{2\gamma_{aa}^{1/2}(
\tau)}\right)\,\right]
\end{eqnarray}
with $Q_{bb}^0$ given by (\ref{eq29}) and 
where $\gamma_{aa}(\tau)=G_{aa}(0)-G_{aa}(\tau)$
\cite{NoteGold}.
We have approximated the sum over $q_x$ by 
an integral, a procedure which becomes
exact in the thermodynamic limit,
and $\mbox{Erfc}(x)$ is the complementary
error function. 
These equations constitute a closed non-linear
system which has to be solved self-consistently
in the limit $n\rightarrow 0$. We consider
the solution of these equations in this limit in
the next section.
\section{Analysis of the variational 
equations for a 2D dipolar interaction. Hierarchical RSB.}
\label{secE}
One needs to take the limit $n\rightarrow 0$ in 
the self-consistent equations
derived above. This implies dealing 
with the matrix structure of the equations
in this limit.
One cannot do this in general. However, there is a 
particular parametrization of these matrices, due to Parisi
\cite{Parisi80}, which is sufficiently general for our purposes.
This is the so called hierarchical RSB.
The diagonal elements of the 
matrix $Q_{ab}$ (with dimensions $n\times n$ 
where $n$ is arbitrarily large) are taken to be 
all equal to $\tilde{q}$ and the off-diagonal elements
are taken to be equal to $q_0$. $Q_{ab}$ is then partioned
in block-diagonal submatrices, the elements of $Q_{ab}$
outside these blocks keeping the value $q_0$ and the
elements inside the blocks taking a new value $q_1$
(except the diagonal elements which keep their value $\tilde q$). 
The procedure is repeated in an equal form for every diagonal 
block submatrix, the off-diagonal elements of these submatrices keeping
the value $q_1$ and the elements inside 
the diagonal sub-blocks taking the value $q_2$. This
procedure is repeated {\em `ad infinitum'}. 
The multiplication rules obeyed by these
hierarchical matrices are analytically continued 
to $n\rightarrow 0$. The off-diagonal elements of $Q_{ab}$
are then parametrized by a function $q(u), \; u\in [0,1]$ and the
diagonal elements are parametrized by the number $\tilde{q}$.
The constraint (\ref{eq29}) takes the form
\begin{equation}
\tilde{q}=\int_{0}^{1}du\, q(u)
\label{eq40}
\end{equation}
and one can show that the matrices $G_{ab}^m$ are also parametrized
by a function $g^m(u)$ and by $\tilde{g}^m$ which parametrizes
the diagonal elements.
In the thermodynamic limit $L_x=L_y\rightarrow\infty$,
$\omega_m\rightarrow \omega$, 
$\tilde{g}^m\rightarrow \tilde{g}(\omega)$ and 
$g^m(u)\rightarrow g(\omega,u)$ and one gets,
using equations (\ref{eq35}-\ref{eq38}), the following relations,
where $\tilde{g}(\tau)$ ($g(\tau,u)$) is the Fourier transform of
$\tilde{g}(\omega)$ ($g(\omega,u)$),
\begin{eqnarray}
\label{eq41}
\hat\Omega^2&=&
\Omega_0^2-\frac{\tilde{\mu}^2\,\Delta\,n}{\sqrt{2\pi}k_B T}
({\cal B}^2+\tilde{g}(\tau=0))^{-1/2}\\
\label{eq42}
q(u)&=&\frac{\tilde{\mu}^2\Delta}{\sqrt{2\pi}k_B T}
[\,{\cal B}^2+2(\tilde{g}(\tau=0)-g(\tau=0,u))\,]^{-1/2}\\
\label{eq43}
f(\omega)&=&
-e^2\int_0^{\infty}\,d\tau\,
(\,1-\cos(\omega\tau)\,)\,\left[\,\frac{1}{2\pi \gamma(\tau)}
-\frac{\tau}{4\,\sqrt{\pi}\,\gamma^{3/2}(\tau)}\,
e^{\frac{\tau^2}{4\gamma(\tau)}}\,
\mbox{Erfc}\left(\frac{\tau}{2\gamma^{1/2}(\tau)}\right)\,\right]\,,
\end{eqnarray}
and where $\gamma(\tau)=\tilde{g}(\tau=0)-\tilde{g}(\tau)$.
One sees from (\ref{eq41}) that when $n\rightarrow 0$, 
$\hat{\Omega}^2=\Omega_0^2$.
The functions $\tilde{g}(\omega)$ and $g(\omega,u)$ are 
in turn given by
\begin{eqnarray}
\label{eq45}
\tilde{g}(\omega)&=&\frac{k_B T}{\sigma(\omega^2+\Omega_0^2)-f(\omega)}
\left(\,1+\frac{q(0)}{\sigma(\omega^2+\Omega_0^2)-f(\omega)}\right.
\nonumber\\
&&\mbox{}+\left.\int_{0}^{1}\frac{dv}{v^2}
\frac{[q](v)}{\sigma(\omega^2+\Omega_0^2)-f(\omega)+[q](v)}\,\right)\\
\label{eq46}
g(\omega,u)&=&\frac{k_B T}{\sigma(\omega^2+\Omega_0^2)-f(\omega)}
\left(\,\frac{q(0)}{\sigma(\omega^2+\Omega_0^2)-f(\omega)}
+\int_{0}^{u}\frac{dv}{v^2}
\frac{[q](v)}{\sigma(\omega^2+\Omega_0^2)-f(\omega)+[q](v)}\right.
\nonumber\\
&&\mbox{}\left.
+\frac{[q](u)}{u[\sigma(\omega^2+\Omega_0^2)-f(\omega)+[q](u)]}
\,\right),
\end{eqnarray}
where $[q](u)=uq(u)-\int_{0}^{u}dv\,q(v)$. Substituting (\ref{eq45})
and (\ref{eq46}) in equation (\ref{eq42}) and differentiating with
respect to $u$, one obtains the equations
\begin{eqnarray}
\label{eq47}
q'(u)&=&0\\
&&\mbox{or}\nonumber\\
q(u)&=&
\left(\frac{\tilde\mu^2\Delta}{\sqrt{2\pi}(k_B T)^{3/2}}
\right)^{2/3}\left(\int_{-\infty}^{\infty}\frac{d\omega}{2\pi}
\frac{1}{(\sigma(\omega^2+\Omega_0^2)-f(\omega)+[q](u))^2}
\right)^{-1/3}\,.\nonumber
\end{eqnarray}
If the first equation is valid in the whole
interval $u\in[0,1]$, its solution is simply $q(u)=const$,
which is the replica symmetric solution. The second equation
is easily solvable if $f(\omega)=0$ (i.e. $e=0$) and
the solution was discussed by M\'ezard and Parisi \cite{Mezard90}.
Also, one has  
\begin{eqnarray}
\label{eq49} 
\overline{\langle (x(\tau)-x(\tau'))^2\rangle}&=&\lim_{n\rightarrow 0}
\frac{2}{n}\sum_{a}(G_{aa}(0) -G_{aa}(\tau-\tau'))\nonumber\\
&=& 2\,\gamma(\tau-\tau')
\end{eqnarray}
which allows for the computation of $\zeta$ once $\gamma(\tau)$
is know. In the presence of dipolar interactions, equation (\ref{eq43})
depends on $\gamma(\tau)$ and one has to resort
to a numerical approach if one wishes to solve it
self-consistently. However, if one simply takes the
zeroth-order result $f(\omega)\approx -\frac{e^2}{2}\mid\omega\mid$,
coming from the expansion of $f(\omega)$ in powers of
$\gamma(\tau)$, one sees that, apart from the contribution
coming from the line tension $\sigma$, the equations
(\ref{eq45}) to (\ref{eq47}) are identical to the
ones obtained by Hazareesing and M\'ezard \cite{Haza99}
in their study of the roughening properties of a 
liquid-gas interface in a disordered solid.
Also, if one expands the last term of equation 
(\ref{eq21}) to quadratic order in $x(\tau)-x(\tau')$,
one can see more directly that the domain-wall energy
has, in this order of approximation, the same analytical
form as that obtained by Kolomeiskii et al. for a
domain-wall in a ferroelastic material 
and as that used by Hazareesing and M\'ezard 
in the study of the problem referred above (provided
one neglects the contribution of the line tension
in our case).
Therefore, if one takes this approximation, one can,
following Robbins and Joanny 
and also Kolomeiskii et al. \cite{Robbins87,Kolomeiskii88}, 
use a Imry-Ma type of argument \cite{Imry75} to show that
the expected value of $\zeta$ is $1/3$, this behaviour
of $\gamma(\tau)$
being valid for lengths $L$ much larger than the Larkin
length \cite{Larkin70} $\xi$, which is given in our problem by
$\xi\sim\frac{e^4\,{\cal B}^3}{\tilde{\mu}^2\,\Delta}$.
When $L\gg\xi$, we can disregard the line
tension contribution to the action, since this
energy term scales with $L^{-1}$. However,
at short length scales compared to $\xi$, 
this term cannot be disregarded,
since it gives an $\omega^2$ contribution to the
denominator of the Green's function appearing
in equation (\ref{eq45}). This contribution gives
rise to a behaviour $\gamma(\tau)\sim\tau$, at 
length scales $\leq {\cal B}$, as can be seen
from the M\'ezard-Parisi solution \cite{Mezard90,Notelsc}. 
Having in mind such behaviour of 
$\gamma(\tau)$, we can write (\ref{eq43}) as
\begin{eqnarray}
\label{eq50}
f(\omega)&=&
-e^2\int_{0}^{\xi}\,d\tau\,
(\,1-\cos(\omega\tau)\,)\,\left[\,\frac{1}{2\pi \gamma(\tau)}
-\frac{\tau}{4\,\sqrt{\pi}\,\gamma^{3/2}(\tau)}\,
e^{\frac{\tau^2}{4\gamma(\tau)}}\,
\mbox{Erfc}\left(\frac{\tau}{2\gamma^{1/2}(\tau)}\right)\,\right]
\nonumber\\
&&\mbox{}-e^2\int_{\xi}^{\infty}\,d\tau\,
(\,1-\cos(\omega\tau)\,)\,\left[\,\frac{1}{2\pi \gamma(\tau)}
-\frac{\tau}{4\,\sqrt{\pi}\,\gamma^{3/2}(\tau)}\,
e^{\frac{\tau^2}{4\gamma(\tau)}}\,
\mbox{Erfc}\left(\frac{\tau}{2\gamma^{1/2}(\tau)}\right)\,\right]\,.
\end{eqnarray}
The integrand in the first term is regular at small $\tau$ 
(due to the behaviour of $\gamma(\tau)$ at small $\tau$) and
its contribution is unimportant if $\omega\,\xi\ll 1$,
which is the set of wave-vectors which determines the
roughening exponent $\zeta$. In the second term, we can use,
if ${\cal B}\ll\xi$,
the asymptotic expansion for $\mbox{Erfc}(x)$. We obtain, 
omiting the regular contribution of the first term,
\begin{equation}
\label{eq51}
f(\omega)\approx
-\frac{e^2}{\pi}\int_{\xi}^{\infty}\,d\tau\,
(\,1-\cos(\omega\tau)\,)\,\left[\,\frac{1}{\tau^2}
-\frac{6\gamma(\tau)}{\tau^4}+\ldots\,
\right]
\end{equation}
where the dots stand for 
terms of order $\gamma^n(\tau)/\tau^{2n+2}$,
with $n\geq 2$. Now, since $\gamma(\tau)\sim\tau^{2\zeta}$ with
$\zeta<1$, then the second term in (\ref{eq51}) is much smaller
than the first one and the omitted terms are of higher order.
We can therefore keep only the first term and we obtain
$f(\omega)\approx -\frac{e^2}{2}\mid\omega\mid$, i.e. the
zeroth order approximation for $f(\omega)$ is a good one. 
The validity of this
approximation of course relies on the fact that $\zeta<1$,
but, as we shall show below, the result obtained
indeed coincides with the Imry-Ma argument, i.e.
$\zeta=1/3$.
If we now substitute this result for $f(\omega)$
in equation (\ref{eq47}) we can perform the integral
(with $\Omega_0=0$) and obtain a closed equation 
for $q(u)$. This equation reads
\begin{equation}
\label{eq52}
q(u)=\left(\frac{\tilde{\mu}^4\Delta^2\sigma^{1/2}}{(k_B T)^3}
\right)^{1/3}\,r^{1/2}(u)\,\left[\,\frac{e^2}{4\sqrt{\sigma}}\,
\frac{r^{1/2}(u)}{[q](u)}
-\mbox{arctanh}\,\left(\
\frac{4\,\sqrt{r(u)\,\sigma}}{e^2}\right)\,\right]^{-1/3}\,,
\end{equation}
where $r(u)=\frac{e^4}{16\sigma}-[q](u)$. Notice that $r(u)$
is positive at small $u$ since $[q](u)\rightarrow 0$ when 
$u\rightarrow 0$. If $r(u)$ becomes negative, this
equation is still valid, provided that we write
$r^{1/2}(u)=i\mid r(u)\mid^{1/2}$ and analytically
continue (\ref{eq52}) to imaginary values. This equation
relates $[q](u)$ to $q(u)$, with $[q]'(u)=u\,q(u)$ by
definition, through a transcendental function and
cannot be easily solved numerically.
It reduces to the equation discussed by Mez\'ard and Parisi
\cite{Mezard90}
if one takes the value of the dipolar interaction to zero 
(i.e. $e^2=0$) and to the equation discussed by 
Hazareesing and Mez\'ard \cite{Haza99}
if one takes the line tension to zero 
(i.e. $\sigma=0$). In these two cases,
this equation reduces to trivial algebraic equations,
which can be easily solved.
In the first case, one obtains $\zeta=1$, which coincides
with the exact result of Zhang \cite{Zhang86} and in the
second case one obtains $\zeta=1/3$, as pointed out above.
In our case, we can nevertheless obtain information regarding
the behaviour of $q(u)$ at small $u$ directly 
from (\ref{eq52}). Performing some algebraic
manipulations with (\ref{eq52}) and using
the identity $\mbox{arctanh}(x)=\log\left(\frac{1-x}{1+x}\right)$,
one can show that
\begin{equation}
\label{eq53}
\lim_{u\rightarrow 0}\,[q](u)\,q^{-3}(u)=
\frac{4(k_B T)^3}{\tilde{\mu}^4\,\Delta^2\,e^2}\,,
\end{equation} 
which shows that, since $[q](u)\rightarrow 0$ when 
$u\rightarrow0$, then one must have $q(u)=A\,u^{1/2}+O(u^{1/2})$
for small $u$ where $O(u^{1/2})$ indicates terms of order
higher than $1/2$ and $A=\left(\frac{\tilde{\mu}^4
\,\Delta^2\,e^2}{12(k_B T)^3}\right)^{1/2}$. Likewise, one
must have $[q](u)=\frac{1}{3}\,A\,u^{3/2}+O(u^{3/2})$,
where $O(u^{3/2})$ indicates terms of order higher
than $3/2$. This coincides with the result obtained
by Hazareesing and Mez\'ard \cite{Haza99}, although their solution
is valid in a finite interval around $0$.
In order to obtain a differential equation
valid $\forall u\in [0,1]$,
we differentiate equation (\ref{eq52})
with respect to $u$. We get
\begin{eqnarray}
\label{eq54}
q'(u)&=&0\\
&&\mbox{or}\nonumber\\
r(u)&=&-\,\frac{1}{2}\,u\,q(u)\nonumber\\
&&\mbox{}+\frac{e^2\,u\,
r^2(u)}{12\,\sqrt{\sigma}\,[q]^2(u)}\,
\left(\frac{\tilde{\mu}^4\Delta^2\sigma^{1/2}}{(k_B T)^3}
\right)^{1/3}\,\left[\,\frac{e^2}{4\sqrt{\sigma}}\,
\frac{r^{1/2}(u)}{[q](u)}
-\mbox{arctanh}\,\left(\
\frac{4\,\sqrt{r(u)\,\sigma}}{e^2}\right)\,\right]^{-4/3}.
\nonumber
\end{eqnarray}
Now, the second equation can, with the aid
of (\ref{eq52}), be written in the following form
\begin{equation}
\label{eq55}
[q]^3(u)=\nu\,[q]^2(u)+\frac{1}{2}\,u\,q(u)
\,(\,[q]^2(u)-\eta\,q^3(u)\,)\,,
\end{equation}
with $\nu\equiv\frac{e^4}{16\,\sigma}$ and 
$\eta\equiv\frac{e^2\,(k_B T)^3}{6\,\tilde{\mu}^4\,\Delta^2\,\sigma}$.
This equation is also valid when $r(u)<0$ without the need
to perform an analytic continuation, which is an advantage
with regard to (\ref{eq52}).
Now, the physical dimensions of $\nu$ and $\eta$ are, according
to these definitions, given by $[\nu]=\mbox{Jm$^{-3}$}$ and 
$[\eta]=\mbox{J$^{-1}$m$^3$}$ (in MKS units).
Furthermore, since $S_0$ has the dimensions of an energy
and both $x(\tau)$ and $\tau$ have the dimensions of
a length, it is also easy to see that the dimensions
of $q(u)$ are the same as those of $\nu$. If we
define $z(u)=\int_0^u\,dv\,q(v)$, one has $q(u)=\frac{dz}{du}$
and $[q](u)=u\,\frac{dz}{du}-z(u)$. Substituting these 
equations in (\ref{eq55}), one obtains a differential
algebraic equation for $z(u)$. This equation reads
\begin{equation}
\label{eq56}
\eta\,u\,\left(\frac{dz}{du}\right)^4+
\left(\,u\,\frac{dz}{du}-z\,\right)^2\,
\left(\,u\,\frac{dz}{du}-2\,(z+\nu)\,\right)=0\,.
\end{equation}
Since $z(u)$ and $\nu$ have the same dimensions,
it is useful to write $z(u)$ in terms of a dimensionless
function $z(u)=\nu\,x^2\,l(x)$, where 
$x=\frac{4\,u^{3/4}}{3\,\alpha^{1/4}}$ 
is a scaling variable with
$\alpha=\eta\,\nu=\frac{e^6(k_B T)^3}{96\,
\tilde{\mu}^4\,\Delta^2\sigma^2}$ being a dimensionless
constant \cite{Note3}. In terms of $l(x)$, the condition (\ref{eq53})
becomes $l(0)=1/2^{5/2}$, since the
$x^2$ power corresponds to a $u^{3/2}$
behaviour at small $u$. Substituting $z(u)$
in terms of $l(x)$ in (\ref{eq56}) we obtain
the following differential algebraic equation
for $l(x)$
\begin{equation}
\label{eq57}
(x\,l'(x)+2\,l(x))^4+\left(\,\frac{3}{4}\,x\,l'(x)+\frac{1}{2}
\,l(x)\,\right)^2\,\left(\,\frac{3}{4}\,x^3\,l'(x)-\frac{1}{2}\,x^2
\,l(x)\,-2\right)=0\,,
\end{equation}
which does not contain any dimensional parameters. The advantage
of this equation with regard to (\ref{eq55}) or (\ref{eq56})
is its scaling form, which means that once we have obtained
the scaling function $l(x)$, 
one simply has to substitute $x$ by its expression
in terms of $u$ to obtain $z(u)$ for all $\nu$ and $\alpha$.
Furthermore, since we have to solve (\ref{eq56}) for
$u\in[0,1]$, this means that we need to solve (\ref{eq57})
for $x\in[0,\frac{4}{3\,\alpha^{1/4}}]$. When $\alpha$
is large \cite{Noteap}, this interval becomes a 
small interval around $0$ and we
can linearize (\ref{eq57}), by writing 
$l(x)=\frac{1}{2^{5/2}}+m(x)$. 
Substituting this equation above and keeping only terms linear
in $m(x)$ and $m'(x)$, we obtain the differential equation
\begin{equation}
\label{eq58}
-\left(\,\frac{x}{8\sqrt{2}}+\frac{3x^3}{512}\,\right)\,m'(x)
+\left(\,\frac{1}{4\sqrt{2}}-\frac{3x^2}{256}\,\right)\,m(x)=
\frac{x^2}{1024\sqrt{2}}\,,
\end{equation}
with solution
\begin{equation}
\label{eq59}
m(x)=\frac{8
Bx^2-128x^2\log x-3\sqrt{2}x^4}{8(32\sqrt{2}+3x^2)^2}\,,
\end{equation}
where $B$ is an arbitrary constant. 
Since the solution $q'(u)=0$ can be valid
in a subinterval of $[0,1]$,
the general solution of (\ref{eq42}) is
\begin{equation}
\label{eq60}
q(u)=
\left\{
\mbox{
\begin{tabular}{c}   
$z'(u)$ if\; $u < u_c$\\
$z'(u_c)$  if\; $u\geq u_c$
\end{tabular}
}
\right.
\end{equation}
where $z(u)=\nu\,x^2\,l(x)$ with $l(x)$ being given by the full
solution of (\ref{eq57}) in the general case or simply by the linear
approximation $l(x)=\frac{1}{2^{5/2}}+\frac{8 Bx^2-128x^2\log x -
3\sqrt{2}x^4}{8(32\sqrt{2}+3x^2)^2}$ in the case of large
$\alpha$. The values of the arbitrary integration constant $B$
appearing in $l(x)$ and of $u_c$ can be determined by substituting the
expression for $q(u)$ obtained from $z(u)$ in the equations
(\ref{eq42}) and (\ref{eq52}). We have solved equation (\ref{eq57})
numerically using the MANPAK algorithm \cite{Rheinboldt96}, 
available at netlib.org.
A comparison between a polynomial fit of the numerical solution
and the linearized solution (\ref{eq59}) is shown in figures 1 and 2,
which shows the correctness of this solution at small $x$.
However, even in the case of large
$\alpha$, where the linear approximation can be used, we still need to
solve the system composed by the transcendental equations
(\ref{eq42}) and (\ref{eq52}),
and one would have to resort to a numerical approach. 
Nevertheless, since the behaviour of $\gamma(\tau)$ at large length 
scales is solely determined
by the behaviour of $[q](u)$ at small $u$, it is still possible to
show, following Hazareesing and M\'ezard
\cite{Haza99}, that the roughening exponent
$\zeta=1/3$. In order to do that, one has to notice that
$[q](u)=1/2\nu x^2l(x)+3/4\nu x^3l'(x)$ for $u\leq u_c$,
$[q](u)=[q](u_c)$ for $u>u_c$.  Substituting this result for $[q](u)$
in equation (\ref{eq45}), we obtain for small $\omega$, using
$l(0)=1/2^{5/2}$ and $l'(0)=0$, which follow from the linear
approximation, the result
\begin{equation}
\label{eq61}
\tilde{g}(\omega)=C[\,\sigma\omega^2+(e^2/2)\mid\omega\mid\,]^{-5/3}+
\ldots
\end{equation}
where the dots indicate terms which diverge less strongly at small
$\omega$ and where $C$ is a numerical constant. From this result
and from the definition of $\gamma(\tau)$,
one immediately concludes that $\zeta=1/3$, in agreement with the
Imry-Ma argument of Robbins and Joanny. 
The determination of the crossover behaviour of $\gamma(\tau)$,
at length scales comparable with $\xi$ as well as the determination
of $\xi$ itself (the Imry-Ma argument just gives the
order of magnitude) requires, as stated above, the solution
of the system of transcendental equations 
(\ref{eq42}) and (\ref{eq52}).
\section{Conclusions}
\label{secF}
We have obtained an expression for the elastic energy 
of a N\'eel domain-wall in a thin ferromagnetic film 
in the presence of dipolar interactions and 
a quenched random field, beyond the quadratic approximation 
for the dipolar energy.
Using the replica trick and a variational ansatz, we have obtained
a set of self-consistent equations for the Green's functions 
of the displacement field of the domain wall.
These equations were solved analytically in the
case of two-dimensional dipolar interactions by making a
quadratic approximation for the dipolar energy,
which was justified on the basis of
the different behavior of the domain-wall at different
length scales. 
The problem is then analogous to the
one of a domain wall in a ferroelastic as studied
by Kolomeiskii et al. and to the
one of a liquid-gas contact line in a disordered solid, as studied
by Hazareesing and M\'ezard. We therefore obtain a value
$\zeta=1/3$ for the roughening exponent of the domain-wall. 

From these calculations, we have obtained some important results.
Firstly, we were able to represent the dipolar interaction in the
domain-wall Hamiltonian as the interaction of the quantum particle
with a annealed gauge field. Also, in 1+1 dimensions, one can
integrate out such a field, leaving us with polaron like quantum
Hamiltonian. Secondly, we have shown that a
generalized replica symmetry breaking ansatz
allows for the treatment of this problem.
Note that this generalization leads to relatively
simple equations in our case, since the non-diagonal
part of the self-energy in replica space is still wave-vector
independent which is due to the short-range correlated 
nature of the random field. The treatment of long-range correlated
disorder is, on the other hand, a much more complicated problem \cite{QRP}. 
Finally, we were able to justify that taking 
the quadratic approximation for two-dimensional dipolar 
interactions is sufficient to obtain
the correct value of the roughening exponent within this approximation. 
\\ \\
{\bf Acknowledgements:}
We acknowledge helpful dis\-cus\-sions 
with  M. M\'ezard, T. Emig and C. Bracher. 
J.E.S. acknowledges financial support during the 
different stages of this work from the EU in the
framework of the Contract ERB-FMBI-CT 97-2816 and the DFG in the
framework of the Sonderforschungsbereich SFB 413/TP C6. 
E. F. acknowledges the 
support of the Deuts\-che Forschungsgemeinschaft 
through an Heisenberg fellowship, contract no. FR850/3.
F. S. acknowledges the support of the 
Deuts\-che Fors\-chungs\-gemein\-schaft Einzelprojekt
Schw. 348/10-1 and of the 
BMBF Verbundprojekt 03-SC5-TUM0.

\newpage
\appendix
\section*{Coupling constants and their physical units}
\label{apA}
Here we collect the definitions of the different coupling
constants used in this paper, together with their dimensions
in the MKS system, denoted by $[\;]$. One uses
the units Joule (J=Kgm$^2$s$^{-2}$), Ampere (A) and 
meter (m).  The `fundamental' physical
constants are the magnetic permitivity of the vacuum $\mu_0$,
the Bohr magneton $\mu_B$, the Land\'e factor for the system
$g_L$, the numerical value of the spin $S$,
the strength of the exchange interaction between nearest
neighbours $J$ (given in J),
the amplitude of the random field
correlations $\Delta$ (given in A$^2$m$^{-2}$)
and the lattice constant $a$. 
Furthermore, in the case
of a SC/FM/SC heterostructure, one also
needs the London penetration depth of the superconductor
$\lambda_L$ and its relative permitivity $\mu_r$.
From the units of these
quantities in the MKS \cite{PhE}, one then determines the dimension
of the coupling constants appearing below. 
\begin{eqnarray}
\sigma&=&\frac{J\,\pi^2\, S^2}{a}\;\;\;\;[\sigma]=\mbox{Jm$^{-1}$},
\label{A.1}
\\
\tilde{\mu}&=&\frac{\mu_0\,\mu_B\,g_L\,\pi\,
S}{a}\;\;\;\;[\tilde{\mu}]=\mbox{JA$^{-1}$},
\\
e^2_{2d}&=&\frac{\mu_0\,\mu_B^2 \,g_L^2\,\pi^2\,S^2}{2\mu_r\,\lambda_L\,a^4}
\;\;\;\;[e^2_{2d}]=\mbox{Jm$^{-2}$},
\\
e^2_{3d}&=&\frac{\mu_0\,\mu_B^2\, g_L^2\,\pi^2\,S^2}{2a^4}
\;\;\;\;[e^2_{3d}]=\mbox{Jm$^{-1}$},
\\
\nu&=&\frac{e^4_{2d}}{16\,\sigma}\;\;\;\;[\nu]=\mbox{Jm$^{-3}$}, 
\\
\eta&=&\frac{e^2_{2d}\,(k_B T)^3}{6\,\tilde{\mu}^4\,\Delta^2\,\sigma}
\;\;\;\;[\eta]=\mbox{J$^{-1}$m$^3$},
\end{eqnarray}
where $e^2_{2d}$ is the value of the coupling constant for 
two-dimensional dipolar interactions (the case treated above)
and $e^2_{3d}$ is the value 
for three dimensional dipolar interactions.
\newpage
\noindent{{\bf  Figure Captions}\\
{\bf Figure 1.} Linearized (continous line) and numerical
(dashed line) solutions of equation \ref{eq57} for the
same initial (and arbitrary) condition $l(1)=0.18095$.
In the case of the numerical solution, the MANPAK
algorithm was iterated down to $x=0$.
It is seen that the two solutions converge to one another
close to $x=0$.\\
{\bf Figure 2.}}    
Left hand side of equation \ref{eq57} evaluated for the
linearized (continous line) and numerical (dashed line)
solutions. It is seen that the linearized solution gives
better results close to $x=0$ but that it deviates
significantly for larger values. The numerical solution
is seen to oscilate around zero. Note that the scale
of the plot is $10^{-6}$.
\newpage
\begin{figure}[htbp]
\psfrag{x}{$x$}
\psfrag{y}{$l(x)$}
\centerline{\epsfxsize 0.75\columnwidth \epsfbox{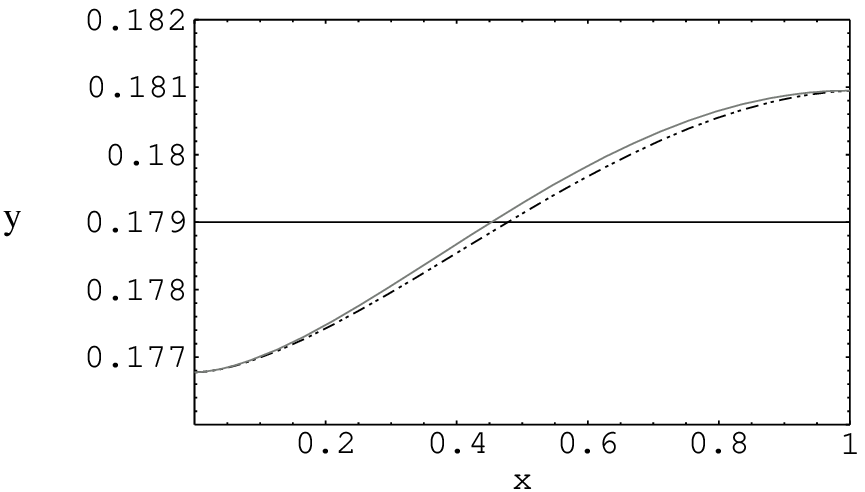}}
\vspace{3mm}
\caption{\label{F1}}
\end{figure}
\newpage
\begin{figure}[htbp]
\psfrag{x}{$x$}
\psfrag{a}{\tiny $4\times 10^{-6}$}
\psfrag{b}{\tiny $3\times 10^{-6}$}
\psfrag{c}{\tiny $2\times 10^{-6}$}
\psfrag{d}{\tiny $1\times 10^{-6}$}
\psfrag{e}{\tiny $0$}
\psfrag{f}{\tiny $-1\times 10^{-6}$}
\psfrag{g}{\tiny $-2\times 10^{-6}$}
\psfrag{h}{\tiny $-3\times 10^{-6}$}
\centerline{\epsfxsize 0.75\columnwidth \epsfbox{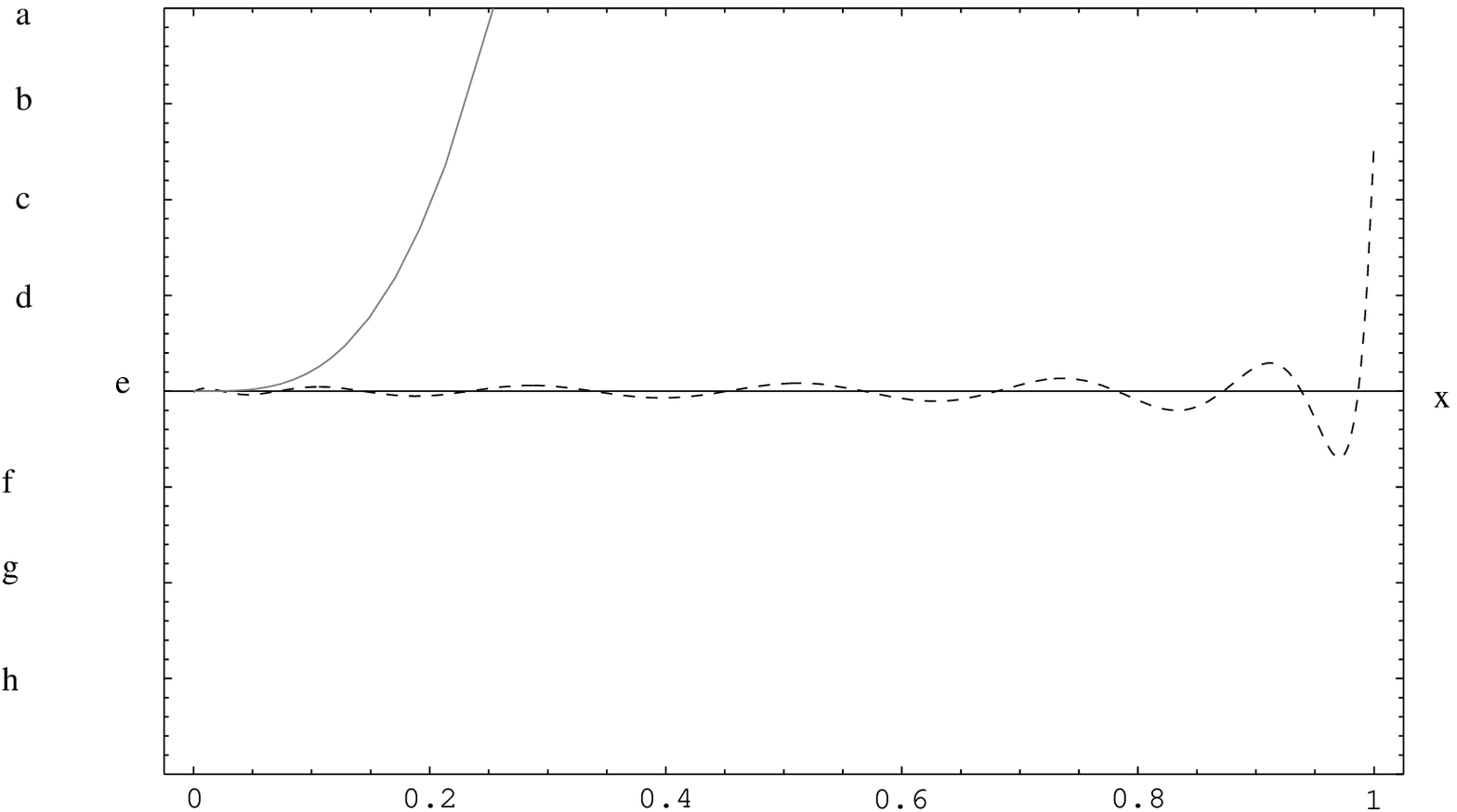}}
\vspace{3mm}
\caption{\label{F2}}
\end{figure}

\begin{thebibliography}{}
\bibitem{Grinstein82} G. Grinstein and S.-K. Ma, Phys. Rev. Lett.
{\bf 49}, 685 (1982).
\bibitem{Villain82} J. Villain, J. Phys. Lett. {\bf 43}, L551 (1982).
\bibitem{Imry75} Y. Imry and S. -K. Ma, Phys. Rev. Lett. {\bf 35},
1399 (1975).
\bibitem{Burgers74} J. M. Burgers, {\it The Nonlinear Diffusion
Equation} (Reidel, Boston, 1974).
\bibitem{Huse85} D. A. Huse, C. L. Henley 
and D. S. Fisher, Phys. Rev. Lett.
{\bf 55}, 2924 (1985).
\bibitem{Zhang86} Y. C. Zhang, J. Phys. A {\bf 19}, L941 (1986).
\bibitem{Mezard90} M. M\'ezard and G. Parisi, J. Phys. A
{\bf 23}, L1229 (1990); J. Phys. I {\bf 1}, 809 (1991).
\bibitem{Gold00} Y. Y. Goldschmidt, Phys. Rev. E {\bf 53}, 343 (1996);
Phys. Rev. B {\bf 56}, 2800 (1997);
Phys. Rev. E {\bf 61}, 1729 (2000).
See also Y. Shiferaw and Y. Y. Goldschmidt, J. Phys. A {\bf 33}, 4461 (2000);
Phys. Rev. E {\bf 63}, 051803 (2001). 
\bibitem{Haza99} A. Hazareesing and M. M\'ezard, Phys. Rev. E
{\bf 60}, 1269 (1999).
\bibitem{Balents93} L. Balents and D. S. Fisher, Phys. Rev. B {\bf 48},
5949 (1993).
\bibitem{Emig98} T. Emig and T. Nattermann, Phys. Rev. Lett. {\bf 81},
1469 (1998); A. Hazareesing and J.-P. Bouchaud, {\em ibidem}, 5953;
T. Emig and T. Nattermann, {\em ibidem}, 5954.
\bibitem{Kolomeiskii88} See also E. B. Kolomeiskii, A. P. Levanyuk and
S. A. Minyukov, Sov. Phys. Solid State {\bf 30}, 311 (1988). 
In this paper, the authors present an extensive 
study of two cases, the one of a 
domain wall in a bulk ferroelectric material in the presence of  
dipolar interactions and the one of a domain wall in a
ferroelastic material, in arbitrary dimensions. 
In the first case, they have concluded, based on the
the Imry-Ma argument given in references 
\cite{Grinstein82,Villain82} and on the change of 
the effective dimension of the problem due to the
presence of dipolar interactions, that the  
value of the roughening exponent is $1/2$, a result 
which also follows from the calculation of Emig and
Nattermann. However, their discussion was,
as stated above, limited to a three dimensional
crystal, i.e. to a domain wall with $D=2$.
The results for the ferroelastic material are also
of interest to the present work, see discussion
in the main text. 
\bibitem{Lajz80} J. Lajzerowicz, Ferroelectrics {\bf 24}, 179 (1980).
\bibitem{Natterman83} T. Natterman, J. Phys. C {\bf 16}, 4125 (1983).
\bibitem{Note0} Note that for a one-dimensional domain-wall the contribution
to the elastic energy coming from the dipolar energy is, in the quadratic
approximation \cite{Natterman83}, $\propto -q^2\log q$ 
at small wave vectors, i.e. the
energy acquires a logarithmic correction. This correction is not
taken into account if one performs the analytical continuation of the
results of Emig and Natterman to $D=1$.
\bibitem{Deker75} U. Deker and F. Haake, Phys. Rev. A {\bf 11}, 2043
(1975).
\bibitem{Feynman56} R. P. Feynman, Phys. Rev. {\bf 97}, 660 (1955).
\bibitem{Spohn86} H. Spohn, J. Phys. A {\bf 19}, 533 (1986).
\bibitem{Parisi80} G. Parisi, J. Phys. A {\bf 13}, 1887 (1980).
\bibitem{Santos00} J. E. Santos, E. Frey
and F. Schwabl, Phys. Rev. B {\bf 63}, 054439 (2001).
\bibitem{NoteMKS}{We shall use the MKS system in this paper since it
allows for an easy analysis of the physical dimensions 
of the quantities involved. See the appendix for
a list of the coupling constants and their physical units in the MKS.}
\bibitem{OA} R. Allespach, J. Mag. Mag. Mat. {\bf 129}, 160 (1994).
\bibitem{Random?} See T. Nattermann, in {\it Spin Glasses and Random
Fields}, A. P. Young (ed.) (World Scientific, Singapore, 1998)
and references therein. 
\bibitem{Notesrf} The rescaling $\tilde{h}^y=h^y/a$ is performed
such that in the continuum limit $a\rightarrow 0$, the correlation
function of the scaled field $\tilde{h}^y$ is a delta function
if we choose $\overline{h_i^\alpha\,h_j^\beta}=
\Delta\,\delta_{ij}\delta^{\alpha\beta}$.
\bibitem{Kleinert?} This mapping appears when 
considering the energy of line-like defects in the
presence of long-range forces. See
H. Kleinert, {\it Gauge Fields in Condensed Matter}, Vol. I
(World Scientific, Singapore, 1989) and references therein.
\bibitem{NoteG} This gauge transformation corresponds to 
adding the total derivative term $\int^{\frac{L_y}{2}}_{-\frac{L_y}{2}}
d\tau\,\frac{d}{d\tau}\,\int_{0}^{x(\tau)}dx'\,
A(x',\tau)$ to the action.
Due to the periodic boundary conditions along the $\tau$ ($y$)
direction the overall contribution of this term is zero.
Also, notice that this gauge
transformation can only be performed in one space dimension.
\bibitem{Edwards75} S. F. Edwards and 
P. W. Anderson, J. Phys. F {\bf 5}, 
965 (1975).
\bibitem{Ada82} J. Adamowski, B. Gerlach and H. Leschke, 
J. Math. Phys. {\bf 23} (2), 243 (1982).
\bibitem{NoteCorr} This form of the correlator allows
for a smooth introduction of a cut-off in the momentum
integrals over $q_x$ and leads naturally to the existence
of a Larkin length. Note that a cut-off is always provided
by the existence of a finite lattice spacing $a$, but its
introduction via the noise correlations makes the equations
analitically more tractable. In the limit in which one 
wants to treat isotropic correlations, one should take
${\cal B}=a$.
\bibitem{NoteGold_1} Note that in our case, we have chosen
the disorder correlator $\Delta(q,\omega)=\Delta(q)$, which implies
that the disorder is locally correlated along the $\tau(y)$ axis.
Goldschmidt has also considered the case $\Delta(q,\omega)
=\Delta(q)\,\delta_{\omega,0}$, which implies that the disorder
correlator is independent of $\tau (y)$. For the corresponding 
problem of a quantum particle in a disordered potential, 
this just means that such potential is independent of time.
\bibitem{NoteGold} In the case of the problem of a quantum particle 
in a static random potential, Goldschmidt has considered an ansatz
in which the only component of $Q_{ab}(\tau -\tau')$  
which has a non-diagonal part is the zero-frequency one, i.e. $Q_{ab}^0$.
\bibitem{Robbins87} M. O. Robbins and J. F. Joanny,
Europhys. Lett. {\bf 3}, 729 (1987).
\bibitem{Larkin70} A. I. Larkin, Sov. Phys. JETP {\bf 31}, 784
(1970). 
\bibitem{Notelsc} Here, we are assuming that the line tension plays
a role at length scales smaller than ${\cal B}$ whereas the dipolar
interaction is unimportant, and that at length scales $\gg \xi$,
the dipolar interaction is relevant whereas the
line tension is not. One should first notice that there is
also a competition between the dipolar term and the line tension term
and that the dipolar term becomes dominant over the 
line tension term at length scales of the
order $\sim \sigma/e^2$ (see Appendix for units). 
Therefore, one must 
have that ${\cal B}\leq \sigma/e^2\leq \xi=\frac{e^4
{\cal B}^3}{\tilde{\mu}^2\Delta}$,
in order for the above assumption to be true, which 
of course depends on the values of ${\cal B}$ and $\Delta$
we choose. If such condition holds, then one 
expects three distinct regimes. 
One regime for which $\mid \tau -\tau'\mid\leq {\cal B}$, 
in which thermal fluctuations are dominant since the 
random field does not play a role at these
length scales. Here the roughening exponent is simply the thermal
one, i.e. $\zeta=1/2$, due to the line tension. A second regime 
${\cal B}\leq \mid \tau -\tau'\mid\leq \xi$, in which one can use a linear
approximation for the random-field and in which the replica symmetric
solution holds. The value of $\zeta$ is
also equal to $1/2$ in this regime, which follows from the analysis
of Hazareesing and M\'ezard. Finally, if $\tau-\tau \geq \xi$,
$\zeta=1/3$, since at these length scales
the line-tension does not play a role and the
replica symmetry breaking solution of Hazareesing 
and M\'ezard holds.
\bibitem{Note3} Obviously, these results are valid provided that
$\nu$ and $\eta$ are not zero (i.e. $e\neq 0$). If $e=0$ then
one obtains from (\ref{eq55}) the solution of M\'ezard and Parisi
\cite{Mezard90}.
\bibitem{Noteap} $\alpha$ large means either that the dipolar
interaction is strong or that the temperature is high or that
either the line tension $\sigma$ or the noise strength $\Delta$
are small.
\bibitem{Rheinboldt96} W. C. Rheinboldt, Computers Math. Applic.
{\bf 32}, 15 (1996); {\em ibidem} {\bf 33}, 31 (1997).
\bibitem{QRP} As stated above, one very important example of 
long-range correlated disorder is the one considered by Goldschmidt
which has a direct application to the problem of a quantum particle
in a static random potential and to the related problem of a polymer
in a medium with long-range correlated disorder. 
Since Goldschmidt has considered a rather general 
variational ansatz where the diagonal part of the self-energy 
is dependent on $\omega$, as we have done above,
such variational ansatz could in principle be used to treat the
problem of a quantum polaron in a random static potential. We thank M.
M\'ezard for calling our attention to this point. 
\bibitem{PhE} See for example D. Halliday, R. Resnick and J. Walker,
{\it Fundamentals of Physics} (John Wiley and Sons, New York, 1993).
\end{thebibliography}
\end{document}